\documentclass[twocolumn,prc,aps,floatfix,showpacs]{revtex4}
\usepackage[dvips]{graphicx}
\newcommand{\mpi}{m_\pi}
\newcommand{\fpi}{f_\pi}
\newcommand{\mKchi}{\tilde{m}_K}
\newcommand{\mK}{m_K}
\newcommand{\wk}{\omega_k}

\newcommand{\gev}{\, {\rm GeV}}
\newcommand{\fm}{\, {\rm fm}}
\newcommand{\non}{\nonumber}

\newcommand{\chiC}{{\cal C}}
\newcommand{\NM}{\mu_N}
\newcommand{\delval}[1]{\frac{\delta #1}{#1}}
\begin{document}
\title{Limits on the temporal variation of the fine structure constant,
quark masses and strong interaction from quasar absorption spectra
and atomic clock experiments}
\author{V.~V.~Flambaum}
\affiliation{ Institute for Advanced Study, Einstein drive,
Princeton, NJ 08540, USA}
\affiliation{
School of Physics, The University of New South Wales, Sydney NSW
2052, Australia}
\author{D.~B.~Leinweber}
\author{A.~W.~Thomas}
\author{R.~D.~Young}
\affiliation{    Special Research Centre for the
		 Subatomic Structure of Matter,
		 and Department of Physics,
         University of Adelaide, Adelaide SA 5005, Australia}
\date{\today}
\begin{abstract}
We perform calculations of the dependence of nuclear magnetic moments
on quark masses and
obtain limits on the variation of
$(m_q/\Lambda_{QCD})$
{}from recent measurements of hydrogen hyperfine (21 cm) and molecular
rotational transitions in quasar absorption systems,
atomic clock
experiments with hyperfine transitions in H, Rb, Cs, Yb$^+$, Hg$^+$
and optical transition in Hg$^+$. Experiments  with
Cd$^+$,  deuterium/hydrogen, molecular SF$_6$ and  
Zeeman transitions in $^3$He/Xe are also discussed.
\end{abstract}


\maketitle

\section{Introduction}
 Interest in the temporal and spatial variation of  major constants of physics
has been recently revived by
 astronomical data which seem to suggest a variation
 of the electromagnetic constant
 $\alpha=e^2/\hbar c$   at the $10^{-5}$ level
 for the time scale 10 billion years, see \cite{alpha}
 (a discussion of
other limits can be found in the review \cite{uzan} and references therein).
 However, an independent experimental confirmation is needed.

The hypothetical unification of all interactions implies that variation
of the electromagnetic interaction constant $\alpha$ should be accompanied
by the variation of masses and the strong interaction constant.
Specific predictions need a model. For example, the grand unification
model discussed in Ref.~\cite{Langacker:2001td} predicts that
the quantum chromodynamic (QCD) scale  $\Lambda_{QCD}$
(defined as the position of the Landau pole in the logarithm for the
running strong coupling constant)
is modified as follows: $\delta \Lambda_{QCD} / \Lambda_{QCD}
\approx 34 \, \delta \alpha / \alpha$.
The variation of quark and electron masses in this model is  given by
$\delta m / m \sim 70 \, \delta \alpha / \alpha $.
This gives an estimate for the variation of the dimensionless ratio
\begin{equation} 
\label{mQCD}
{\delta(m/ \Lambda_{QCD}) \over(m/\Lambda_{QCD})} \sim \, 35 {\delta \alpha
\over \alpha}
\end{equation}
This result is strongly model-dependent (for example, 
the coefficient may be an order of magnitude smaller and
 even of opposite sign \cite{dent}).
However, the large coefficients in these expressions are generic for
grand unification models, in which modifications come from high energy scales:
they appear because the running strong coupling constant and
Higgs constants (related to mass) run faster than $\alpha$.
This means that if these models
are correct the variation
of masses and the strong interaction scale 
may be easier to detect than the variation
of $\alpha$.

One can only measure the variation of 
dimensionless quantities and therefore we want to extract from the measurements
the variation
of the dimensionless ratio $m_q/\Lambda_{QCD}$ -- where $m_q$ is the quark
mass (with the dependence on the renormalization point removed).
A number of limits on the variation of $m_q/\Lambda_{QCD}$
have been obtained recently from consideration
of Big Bang Nucleosynthesis, quasar absorption spectra
and the Oklo natural nuclear reactor, which was active about
1.8 billion years ago \cite{FS,oliv,dmitriev,FS1} (see also
\cite{Murphy1,Cowie,Oklo,c12,savage}). Below we consider
the limits which follow from  quasar absorption radio spectra and
laboratory atomic clock comparisons.
Laboratory limits with a time base of the order one year are
especially sensitive to oscillatory variations of
fundamental constants. A  number of relevant measurements
have been performed already and even larger numbers have been started
or are planned. The increase in precision is happening very fast. 

It has been pointed out by Karshenboim \cite{Karschenboim}
that measurements of ratios of hyperfine structure intervals
in different atoms are sensitive to any  
variation of nuclear magnetic moments.
{}First rough estimates of the dependence of nuclear magnetic
moments on 
$m_q/\Lambda_{QCD}$ and limits on the variation of this ratio with time 
were obtained in Ref.~\cite{FS}. Using  
H, Cs and Hg$^+$ measurements \cite{prestage,Cs},
we obtained a limit on the variation of $m_q/\Lambda_{QCD}$
of about $5 \cdot 10^{-13}$ per year.
Below we calculate the dependence of nuclear magnetic moments
on  $m_q/\Lambda_{QCD}$ and  obtain the limits 
{}from recent atomic clock
experiments with hyperfine transitions in H, Rb, Cs,Yb$^+$,Hg$^+$
and the optical transition in Hg$^+$.
It is convenient to assume that the strong interaction scale,  
$\Lambda_{QCD}$, does not vary, so we will speak about the variation
of masses (this means that we measure masses in units
of $\Lambda_{QCD}$).
We shall restore the explicit appearance of $\Lambda_{QCD}$ in 
the final answers.
   
The hyperfine structure constant can be presented in the following
{}form
\begin{equation}\label{A}
A=const \times [\frac{m_e e^4}{\hbar ^2}] [ \alpha ^2 F_{rel}( Z \alpha)]
[\mu \frac{m_e}{m_p}]
\end{equation}
The factor in the first bracket is an atomic unit of energy. The second 
``electromagnetic'' bracket determines the dependence on $\alpha$.
An approximate expression for the relativistic correction factor (Casimir
{}factor) for an s-wave electron is the following
\begin{equation}\label{F}
F_{rel}= \frac{3}{\gamma (4 \gamma^2 -1)} \, , 
\end{equation}
where $\gamma=\sqrt{1-(Z \alpha)^2}$ and Z is the nuclear charge.
Variation of $\alpha$ leads to the following variation of $F_{rel}$
\cite{prestage}:
\begin{equation}
\frac{\delta F_{rel}}{F_{rel}}=K \frac{\delta \alpha}{\alpha} \, ,
\label{dF}
\end{equation}
\begin{equation}\label{K}
K=\frac{(Z \alpha)^2 (12 \gamma^2 -1)}{\gamma^2 (4 \gamma^2 -1)} \, .
\end{equation}
More accurate numerical many-body calculations \cite{dzuba1999}
of the dependence of the hyperfine structure on $\alpha$ have shown
that the coefficient $K$ is slightly larger than that given by this
{}formula. For Cs ($Z$=55) $K$= 0.83 (instead of 0.74),
{}for Rb $K$=0.34
(instead of 0.29) and finally for Hg$^+$
$K$=2.28 (instead of 2.18).

The last bracket in Eq.~(\ref{A})  contains the dimensionless
nuclear magnetic moment $\mu$ 
(i.e., the nuclear magnetic moment $M=\mu\frac{e\hbar}{2 m_p c}$),
electron mass $m_e$ and proton mass $m_p$. We may also include
a small correction arising from the finite nuclear
size. However, its contribution is insignificant.

Recent experiments measured the time dependence of the ratios of
the hyperfine structure intervals of $^{199}$Hg$^+$ and H \cite{prestage},
$^{133}$Cs and $^{87}$Rb \cite{marion} and the ratio of the optical frequency
in Hg$^+$ to the hyperfine frequency of $^{133}$Cs~\cite{bize}.  
In the ratio of two
hyperfine structure constants for different atoms time dependence
may appear from the ratio of the factors $F_{rel}$ (depending on $\alpha$)
as well as from the ratio of nuclear magnetic 
moments (depending on $m_q/\Lambda_{QCD}$).
Magnetic moments in a single-particle approximation (one unpaired nucleon)
are:  
\begin{equation}\label{mu+}
\mu=(g_s + (2 j-1) g_l)/2 \, ,
\end{equation}
for $j=l+1/2$.
\begin{equation}\label{mu-}
\mu=\frac{j}{2(j+1)}(-g_s + (2 j+3) g_l)
\end{equation}
{}for $j=l-1/2$. Here the orbital g-factors are
$g_l=1$ for a valence proton and $g_l=0$ for a valence
neutron. The present values of the spin g-factors, $g_s$, are
$g_p=5.586$ for proton and $g_n=-3.826$ for neutron.
They depend on $m_q/\Lambda_{QCD}$.
The light quark masses are only about $1 \%$ of the nucleon mass
($m_q=(m_u+m_d)/2 \approx$ 5 MeV) and the nucleon magnetic moment remains
{}finite in the chiral  limit, $m_u=m_d=0$. Therefore, one might think that 
the corrections to $g_s$ arising from the finite quark masses 
would be very small.
However, through the mechanism of spontaneous chiral symmetry
breaking, which leads to contributions to hadron properties from
Goldstone boson loops, one may expect some enhancement of the effect of
quark masses~\cite{Leinweber:2001ui}. 
The natural framework for discussing such corrections is
chiral perturbation theory and we discuss these chiral corrections next.

\section{Chiral perturbation theory results for nucleon magnetic
moments and masses}

In recent years there has been tremendous progress in the calculation
of hadron properties using lattice QCD. Moore's Law, in combination
with sophisticated algorithms, means that one can now make extremely
accurate calculations for light quark masses ($m_q$) larger than 50
MeV. However, in order to compare with experimental data, it is still
necessary to extrapolate quite a long way as a function of quark mass.
This extrapolation is rendered non-trivial by the spontaneous breaking
of chiral symmetry in QCD, which leads to Goldstone boson loops and,
as a direct consequence, non-analytic behaviour as a function of quark
mass~\cite{Bernard:2002yk,Thomas:2002sj}.  {}Fortunately the most
important nonanalytic contributions are model independent, providing a
powerful constraint on the extrapolation procedure.

In the past few years the behaviour of hadron properties as a function
of quark mass has been studied over a much wider range than one needs
for the present
purpose~\cite{Thomas:2002sj,Leinweber:1999ig,Leinweber:1998ej,Hemmert:2003ta,Hemmert:2003wt,Bernard:2003rp,Gockeler:2003ay}. 
One can therefore apply the successful
extrapolation formulas developed in the context of lattice QCD with
considerable confidence.

The key qualitative feature learnt from the study of lattice data is
that Goldstone boson loops are strongly suppressed once the Compton
wavelength of the boson is smaller than the source. Inspection of
lattice data for a range of observables, from masses to charge radii
and magnetic moments, reveals that the relevant mass scale for this
transition is $m_q \sim 50$ MeV -- i.e., $m_\pi \sim 400-500$
MeV~\cite{Thomas:2002sj,Detmold:2001hq}.  The challenge of chiral
extrapolation is therefore to incorporate the correct, model
independent non-analytic behaviour dictated by chiral symmetry while
ensuring excellent convergence properties of the chiral expansion in
the large mass region, as well as maintaining the model independence
of the results of the extrapolation.  Considerable study of this
problem has established that the use of a finite range regulator (FRR)
fulfils all of these 
requirements~\cite{Donoghue:1998bs,Leinweber:2003dg,Young:2002ib}. 
Indeed, in the case of the mass of
the nucleon, it has been shown that the extrapolation from $m_\pi^2
\sim 0.25$ GeV$^2$ to the physical pion mass -- a change of $m_q$ by a
factor of 10 -- can be carried out with a systematic error less than
1\%~\cite{Leinweber:2003dg}.  In the following we apply this same
method to calculate the change in the nucleon mass, corresponding to
quark mass changes at the level of 0.1\% or less, as required in the
present context.

\subsection{Variation of the nucleon mass with quark mass}
The expansion for the mass of the nucleon given in
Ref.~\cite{Young:2002ib,Leinweber:2003dg} is:
\begin{equation}
M_N = a_0 + a_2 \mpi^2 + a_4 \mpi^4 + a_6 \mpi^6 + \sigma_{N\pi} +
\sigma_{\Delta\pi} + \sigma_{\rm tad}\, , 
\label{eq:MN}
\end{equation}
where the chiral loops which given rise, respectively, to the leading
and next-to-leading nonanalytic (LNA and NLNA) behaviour are:
\begin{eqnarray}
\sigma_{N\pi}      &=& -\frac{3}{32\,\pi\fpi^2}\, g_A^2\, I_M(m_\pi,
\Delta_{NN}, \Lambda) \\
\sigma_{\Delta\pi} &=& -\frac{3}{32\,\pi\fpi^2}\, \frac{32}{25} g_A^2\,
I_M(m_\pi, \Delta_{N \Delta}, \Lambda) \\
\sigma_{\rm tad} &=& -\frac{3}{16\,\pi^2\fpi^2}\,c_2\mpi^2\, I_T(\mpi,\Lambda)\, ,
\label{eq:sigmas}
\end{eqnarray}
and the relevant integrals are defined (in heavy baryon approximation)  
as:
\begin{eqnarray}
I_M(m_P,\Delta_{BB'},\Lambda) = \frac{2}{\pi}\int_0^\infty dk\,
\frac{k^4 u^2(k,\Lambda)}{\omega_k (\Delta_{BB'} + \omega_k )} \\
I_T(\mpi,\Lambda) =
\int_0^\infty dk \left(\frac{2 k^2 u^2(k)}{\sqrt{k^2+\mpi^2}}\right) -
  t_0
\, , \label{eq:tadint}
\end{eqnarray}
with $\omega_k = \sqrt{k^2+m_P^2}$ and $\Delta_{BB'}$ the relevant
baryon mass difference (i.e., $M_{B'}-M_B$). 
We take the $\Delta$--$N$ mass splitting, $\Delta=M_\Delta - M_N$, to
have its physical value (0.292 GeV), while $g_A = 1.26$. The regulator
function, $u(k,\Lambda)$, is taken to be a dipole with mass
$\Lambda=0.8\gev$. In Eq.~(\ref{eq:tadint}) $t_0$, defined such that
$I_T$ vanishes at $\mpi=0$, is a local counter term introduced in
FRR to ensure a linear relation for the renormalisation of $c_2$.

The model independence of the expansion given in Eq.~(\ref{eq:MN}) is
ensured by fitting the unknown coefficients to the physical nucleon
mass and lattice data from the CP-PACS Collaboration
\cite{AliKhan:2001tx}, yielding:
$a_0=1.22,a_2=1.76,a_4=-0.829,a_6=0.260$ (with all parameters
expressed in the appropriate powers of GeV). With these parameters
fixed one can evaluate the rate of change of the mass of the nucleon
with quark or pion mass at the physical pion mass:
\begin{equation}
m_q\, \frac{\partial}{\partial \, m_q} M_N = \mpi^2\, \frac{\partial}
{\partial \, \mpi^2} M_N = 0.035 \, {\rm GeV} \, ,
\label{eq:sigma-comm}
\end{equation}
a quantity commonly known as the pion-nucleon sigma commutator.
Using Eq.~(\ref{eq:sigma-comm}) one finds the relationship 
(in terms of dimensionless quantities):
\begin{eqnarray}
\frac{\delta M_N}{M_N} &=& \frac{m_\pi^2}{M_N} 
\frac{\partial \, M_N}{\partial \, \mpi^2} \frac{\delta m_q}{m_q} \\
&=& 0.037 \, \frac{\delta m_q}{m_q}
\label{eq:mass-frac}
\end{eqnarray}

The extension of this procedure to the effect of a variation in the
strange quark mass is similar, but one must include the variation
arising from $\eta$-Nucleon loops, as well as Kaon loops with intermediate
$\Sigma$ or $\Lambda$ baryons.
\begin{equation}
\sigma^K_{N\Sigma} + \sigma^K_{N\Lambda} + \sigma^\eta_{NN}
\label{eq:sigma-strange}
\end{equation}
These contributions can be expressed as
\begin{equation}
\sigma^P_{BB'} = - \frac{3}{32\pi\fpi^2} \, G^P_{BB'} \, 
I_M(m_P,\Delta_{BB'},\Lambda)
\end{equation}
with $G^P_{BB'}$ the associated coupling squared. 
Once again we select the dipole regulator:
\begin{equation}
u(k,\Lambda)=\left(\frac{\Lambda^2}{\Lambda^2+k^2}\right)^2
\, .  
\end{equation}

{}For the relevant diagrams, $N\to \Sigma K$, $N\to\Lambda K$ and $N\to
N\eta$, we have
\begin{eqnarray}
G^K_{N\Sigma}  &=& \frac{1}{3}(D-F)^2 \non\\
G^K_{N\Lambda} &=& \frac{1}{9}(3F+D)^2 \non\\
G^\eta_{NN}    &=& \frac{1}{9}(3F-D)^2
\end{eqnarray}
where we take $F = 0.50$ and $D = 0.76$.
We use the Gell Mann-Oakes-Renner relation in the SU(2) chiral limit 
to relate the variation of the kaon mass in the chiral SU(2) limit,
$\mKchi = \sqrt{\mu_K^2 - \frac{1}{2} \mu_\pi^2} = 0.484$ GeV (with
$\mu_{\pi \{K\}}$, the physical pion\{kaon\} mass), to the variation of
the strange quark mass ($\delta \mKchi^2 / \mKchi^2 = \delta m_s/m_s$).
Hence the  
variation of the nucleon mass with strange quark mass is given
by:
\begin{equation}
\frac{\delta M_N}{M_N} = \left\{ \frac{\mKchi^2}{M_N} 
\frac{\partial}{\partial \, \mKchi^2} \left(
\sigma^K_{N\Sigma} + \sigma^K_{N\Lambda} + \sigma^\eta_{NN} \right)
\right\} \frac{\delta m_s}{m_s} \, .
\label{eq:MN-ms-variation}
\end{equation}
Using the dipole regulator mass, $\Lambda=0.8\gev$, 
Eq.~(\ref{eq:MN-ms-variation}) leads to the result:
\begin{equation}
\frac{\delta M_N}{M_N} = 0.011 \, \frac{\delta m_s}{m_s} \, . 
\label{eq:dmds}
\end{equation}

\subsection{Variation of proton and neutron magnetic moments with quark mass}
The treatment of the mass dependence of the nucleon magnetic moments is
very similar to that for the masses. Once again the loops which give
rise to the LNA and NLNA behaviour are evaluated with a FRR, while the
smooth, analytic variation with quark mass is parametrized by fitting 
relevant lattice data with a finite number of adjustable constants.

For the lattice data we use CSSM Lattice Collaboration results
\cite{Zanotti:2003gc} of nucleon 3-point functions. Results are
obtained using established techniques in the extraction of form factor
data \cite{Leinweber:1990dv}. Similar calculations have also been
recently reported by the QCDSF Collaboration~\cite{Gockeler:2003ay}.  
We use the two heaviest
simulation results, $\mpi^2\sim 0.6$--$0.7\gev^2$
\cite{Zanotti:2003gc}. These simulations were performed with the FLIC
fermion action~\cite{Zanotti:2001yb} on a $20^3\times 40$ lattice at
$a=0.128\fm$.

\begin{figure}[tbp]
\begin{center}
{\includegraphics[width=6cm,angle=0]{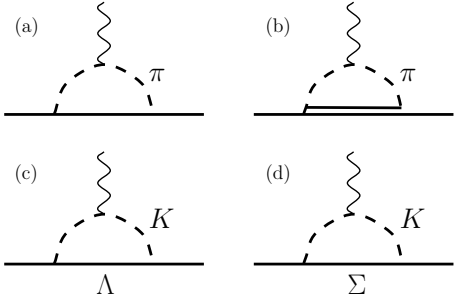}}
\end{center}
\vspace*{-0.5cm}
\caption{Chiral corrections to the nucleon magnetic moments included in
the present work.
}
\label{fig:1}
\end{figure}

In the magnetic moment case the formulae are a little more
complicated, so we leave the details for the Appendix. Suffice it to
say here that the relevant processes are shown in Fig.~\ref{fig:1}.
Again we use a dipole form for the regulator with $\Lambda = 0.8$ GeV.

Having parametrized the neutron and proton magnetic moments as a
function of $m_\pi$, the fractional change versus $m_q$ or $m_s$ is
given by:
\begin{equation}
\delval{\mu} = \left\{ \frac{\mpi^2}{\mu} 
\frac{\partial \,\mu}{\partial \, \mpi^2}
\right\} \delval{m_q}
\end{equation}
\begin{equation}
\delval{\mu} = \left\{ \frac{\mKchi^2}{\mu} 
\frac{\partial \, \mu}{\partial \, \mKchi^2}
\right\} \delval{m_s}\, .
\end{equation}
The numerical results may then be summarised as:
\begin{equation}
\delval{\mu_p} = -0.087 \delval{m_q}
\label{gpsq}
\end{equation}
\begin{equation}
\delval{\mu_p} = -0.013 \delval{m_s}
\label{gpss}
\end{equation}
\begin{equation}
\delval{\mu_n} = -0.118 \delval{m_q}
\label{gnsq}
\end{equation}
\begin{equation}
\delval{\mu_n} = 0.0013 \delval{m_s}
\label{gnss}
\end{equation}
\begin{equation}
\delval{(\mu_p/\mu_n)} = 0.031 \delval{m_q}
\end{equation}
\begin{equation}
\delval{(\mu_p/\mu_n)} = -0.015 \delval{m_s} \, .
\end{equation}
%

\section{Dependence of atomic transition frequencies
on fundamental constants}
Using the results of the previous section we can now use
Eqs.~(\ref{mu+},\ref{mu-}) to study the variation of nuclear magnetic 
moments. For all even Z nuclei with valence
neutron ($^{199}$Hg,$^{171}$Yb,$^{111}$Cd, etc.)
we obtain $\frac{\delta \mu}{\mu} =\frac{\delta g_n}{g_n}$.
{}For $^{133}$Cs we have a valence proton with $j$=7/2, $l$=4 and 
\begin{equation}\label{Cs}
\frac{\delta \mu}{\mu} =
\, 0.110 \frac{\delta m_q}{m_q}+0.017 \, \frac{\delta m_s}{m_s}
\end{equation}
{}For $^{87}$Rb we have valence proton with $j$=3/2, $l$=1 and 
\begin{equation}\label{Rb}
\frac{\delta \mu}{\mu} =
- 0.064 \, \frac{\delta m_q}{m_q}
-0.010 \, \frac{\delta m_s}{m_s}
\end{equation}

As an intermediate result it is convenient to present the dependence of
the ratio of the hyperfine
constant, $A$, to the atomic unit of energy  $E=\frac{m_e e^4}{\hbar ^2}$ (or
the energy of the 1s-2s transition in hydrogen, which is equal to 3/8 $E$)
on a variation of the fundamental constants. We introduce a parameter $V$
defined by the relation
\begin{equation}\label{V}
\frac{\delta V}{V} \equiv \frac{\delta (A/E)}{A/E} \, . 
\end{equation}
We start from the hyperfine structure of $^{133}Cs$
which is used as a frequency standard. Using Eqs.~(\ref{A},\ref{Cs}) 
we obtain 
\begin{equation}\label{VCs}
V(^{133}Cs)=\alpha^{2.83}(\frac{m_q}{\Lambda_{QCD}})^{0.110}
(\frac{m_s}{\Lambda_{QCD}})^{0.017}
\frac{m_e}{m_p}
\end{equation}
The factor $\frac{m_e}{m_p}$ will cancel out in the ratio of hyperfine
transition frequencies. However, it will survive in comparison between
hyperfine and optical or molecular transitions (see below).  According
to Eqs.~(\ref{eq:mass-frac}) and (\ref{eq:dmds}) the relative
variation of the electron to proton mass ratio can be described by the
parameter
\begin{equation}\label{memp}
X(m_e/m_p)=
(\frac{m_q}{\Lambda_{QCD}})^{-0.037}
(\frac{m_s}{\Lambda_{QCD}})^{-0.011}\frac{m_e}{\Lambda_{QCD}}
\end{equation}
which can be substituted into Eq.~(\ref{VCs}) instead of $m_e/m_p$.
This gives an expression which is convenient to use for comparison
with optical and molecular vibrational or rotational transitions
\begin{equation}\label{VCs1}
V(^{133}Cs)=\alpha^{2.83}(\frac{m_q}{\Lambda_{QCD}})^{0.073}
(\frac{m_s}{\Lambda_{QCD}})^{0.006}
\frac{m_e}{\Lambda_{QCD}}
\end{equation}
The dependence on the strange quark mass is relatively weak.
Therefore, it may be convenient to assume that the relative variation
of the strange quark mass is the same as the relative variation
of the light quark masses (this assumption is motivated by the Higgs
mechanism of mass generation) and to use an approximate expression\\
$V(^{133}Cs)\approx\alpha^{2.83}(\frac{m_q}{\Lambda_{QCD}})^{0.13}
\frac{m_e}{m_p}$.
 
{}For hyperfine transition frequencies in other atoms we obtain
\begin{equation}\label{VRb}
V(^{87}Rb)=\alpha^{2.34}(\frac{m_q}{\Lambda_{QCD}})^{-0.064}
(\frac{m_s}{\Lambda_{QCD}})^{-0.010}
\frac{m_e}{m_p}
\end{equation}
\begin{equation}\label{VH}
V(^{1}H)=\alpha^{2}(\frac{m_q}{\Lambda_{QCD}})^{-0.087}
(\frac{m_s}{\Lambda_{QCD}})^{-0.013}
\frac{m_e}{m_p}
\end{equation}
\begin{equation}\label{VD}
V(^{2}H)=\alpha^{2}(\frac{m_q}{\Lambda_{QCD}})^{-0.018}
(\frac{m_s}{\Lambda_{QCD}})^{-0.045}
\frac{m_e}{m_p}
\end{equation}
\begin{equation}\label{Hg}
V(^{199}Hg^+)=\alpha^{4.3}(\frac{m_q}{\Lambda_{QCD}})^{-0.118}
(\frac{m_s}{\Lambda_{QCD}})^{0.0013}
\frac{m_e}{m_p}
\end{equation}
\begin{equation}\label{Yb}
V(^{171}Yb^+)=\alpha^{3.5}
(\frac{m_q}{\Lambda_{QCD}})^{-0.118}
(\frac{m_s}{\Lambda_{QCD}})^{0.0013}\frac{m_e}{m_p}
\end{equation}
\begin{equation}\label{Cd}
V(^{111}Cd^+)=\alpha^{2.6}
(\frac{m_q}{\Lambda_{QCD}})^{-0.118}
(\frac{m_s}{\Lambda_{QCD}})^{0.0013}\frac{m_e}{m_p}
\, .
\end{equation}
Note that the hyperfine frequencies of all even-Z atoms
where the nuclear magnetic moment
is determined by a valence neutron have the same dependence
on quark masses.
 
\section{Limits on  variation of fundamental constants}
Now we can use these results to place limits on the possible variation
of the fundamental constants from particular measurements.
Let us start from the measurements of quasar absorption spectra.
Comparison of the atomic H 21 cm (hyperfine) transition with molecular
rotational transitions \cite{Murphy1} gave limits for the 
variation of $Y_g\equiv\alpha^2 g_p$. In Refs.~\cite{FS,DF}
it was suggested
that one might use these limits
to estimate variation of  $m_q/\Lambda_{QCD}$.
According to Eqs.~(\ref{gpsq}) and (\ref{gpss}) 
the relative variation
of $Y_g$ can be replaced by the relative variation of $Y$
($\delta Y/Y=\delta Y_g/Y_g$)
\begin{equation}\label{cosmic}
Y=\alpha^{2}
(\frac{m_q}{\Lambda_{QCD}})^{-0.087}
(\frac{m_s}{\Lambda_{QCD}})^{-0.013}
\, .
\end{equation}
Then the measurements in Ref.~\cite{Murphy1} lead to the following limits
on the variation of $Y$:\\
 $\delta Y/Y=(-0.20 \pm 0.44) 10^{-5}$ for redshift z=0.2467 and \\
 $\delta Y/Y=(-0.16 \pm 0.54) 10^{-5}$ for  z=0.6847.\\ 
The second limit corresponds to roughly t=6 billion years ago.
There is also a  limit on variation of $X_m\equiv\alpha^2 g_p m_e/m_p$ 
obtained in Ref.~\cite{Cowie}.
This limit was interpreted as a limit on variation of $\alpha$
or $m_e/m_p$.
The relative variation of $X_m$
can be replaced by the relative variation of
\begin{equation}\label{cosmic2}
X=\alpha^{2}
(\frac{m_q}{\Lambda_{QCD}})^{-0.124}
(\frac{m_s}{\Lambda_{QCD}})^{-0.024}\frac{m_e}{\Lambda_{QCD}}\, .
\end{equation}
The dependence on quark masses appears from both the proton g-factor
and the proton mass.
The measurement in Ref.~\cite{Cowie} leads to the following limit on
variation of $X$:\\
$\delta X/X=(0.7\pm 1.1) 10^{-5}$ for z=1.8.

Now let us discuss the limits obtained from the laboratory
measurements of the time dependence of hyperfine structure intervals.
The dependence of the ratio of frequencies $A(^{133}$Cs)/$A(^{87}$Rb)
can be presented in the following form
\begin{eqnarray}
X(Cs/Rb)=\frac{V(Cs)}{V(Rb)} \hspace*{40mm}\non\\
=\alpha^{0.49} [m_q/\Lambda_{QCD}]^{0.174}
[m_s/\Lambda_{QCD}]^{0.027}\, 
\label{CsRb}
\end{eqnarray}
and the result of the measurement in Ref.~\cite{marion} may be presented
as a limit on variation of the parameter $X$:
\begin{equation}\label{limitCsRb}
\frac{1}{X(Cs/Rb)}\frac{dX(Cs/Rb)}{dt}=
(0.2 \pm 7) \times 10^{-16}/year \, .
\end{equation}
Note that if the relation (\ref{mQCD}) 
were correct, the variation of $X(Cs/Rb)$
would be dominated by  variation of $[m_q/\Lambda_{QCD}]$.
The relation (\ref{mQCD}) would give
$X(Cs/Rb) \propto \alpha^{8}$.

For $A(^{133}$Cs)/$A$(H) we have
\begin{eqnarray}
X(Cs/H)=\frac{V(Cs)}{V(H)} \hspace*{40mm}\non\\
=\alpha^{0.83} [m_q/\Lambda_{QCD}]^{0.196}[m_s/\Lambda_{QCD}]^{0.030}\, 
\label{CsH}
\end{eqnarray}
and the result of the measurements in Ref.~\cite{Cs}  may be presented
as
\begin{equation}\label{limitCsH}
 |\frac{1}{X(Cs/H)}\frac{dX(Cs/H)}{dt}|<
 5.5 \times 10^{-14}/year \, . 
\end{equation}

{}For $A(^{199}$Hg)/$A$(H) we have
\begin{eqnarray}
X(Hg/H)=\frac{V(Hg)}{V(H)} \hspace*{40mm}\non\\
\approx
\alpha^{2.3}[m_q/\Lambda_{QCD}]^{-0.031}[m_s/\Lambda_{QCD}]^{0.015}\, .
\label{HgH}
\end{eqnarray}
The result of the measurement in Ref.~\cite{prestage}  may be presented
as
\begin{equation}\label{limitHgH}
 |\frac{1}{X(Hg/H)}\frac{dX(Hg/H)}{dt}|<
 8 \times 10^{-14}/year \, .
\end{equation}
Note that because the dependence on masses and strong interaction
scale is very weak here, this experiment may be interpreted as
a limit on the variation of $\alpha$.

In Ref.~\cite{Karschenboim} a limit was obtained on the variation of the ratio
of hyperfine transition frequencies  $^{171}$Yb$^+$/$^{133}$Cs 
(this limit is based on the measurements of Ref.~\cite{Fisk}).
Using Eqs.~(\ref{VCs},\ref{Yb}) we can present the result as a limit
on $X(Yb/Cs)=\alpha^{0.7}[m_q/\Lambda_{QCD}]^{-0.228}
[m_s/\Lambda_{QCD}]^{-0.015}$:
\begin{equation}\label{limitYbCs}
 \frac{1}{X(Yb/Cs)}\frac{dX(Yb/Cs)}{dt} \approx -1(2)
 \times 10^{-13}/year \, . 
\end{equation}

The optical clock transition energy $E(Hg)$ ($\lambda$=282 nm)
in the Hg$^+$ ion can be presented in the following form:
\begin{equation}\label{E}
E(Hg)=const \times [\frac{m_e e^4}{\hbar ^2}] F_{rel}(Z \alpha) \, . 
\end{equation}
Numerical calculation
of the relative variation
of $E(Hg)$ has given~\cite{dzuba1999}:
\begin{equation}\label{dE}
\frac{\delta E(Hg)}{E(Hg)}=-3.2\frac{\delta \alpha}{\alpha} \, . 
\end{equation}
This corresponds to $V(HgOpt)=\alpha^{-3.2}$. 
Variation of the ratio of the Cs hyperfine splitting
 $A(Cs)$ to this optical transition energy is described by
$X(Opt)=V(Cs)/V(HgOpt)$:
\begin{equation}\label{CsHgE}
X(Opt)=\alpha^{6}(\frac{m_q}{\Lambda_{QCD}})^{0.073}
 (\frac{m_s}{\Lambda_{QCD}})^{0.006} (\frac{m_e}{\Lambda_{QCD}}) \, .
\end{equation}
Here we used Eq.~(\ref{VCs1}) for $V(Cs)$. The work of Ref.~\cite{bize}
gives the limit on variation of this parameter:
\begin{equation}\label{limitCsHgE}
 |\frac{1}{X(Opt)}\frac{dX(Opt)}{dt}|<
 7 \times 10^{-15}/year \, . 
\end{equation}

Molecular vibrational transitions frequencies are proportional
to $(m_e/m_p)^{1/2}$. Based on Eq.~(\ref{memp}) we
may describe the relative variation of vibrational frequencies 
by the parameter
\begin{equation}\label{vib}
V(vib)=
(\frac{m_q}{\Lambda_{QCD}})^{-0.018}
(\frac{m_s}{\Lambda_{QCD}})^{-0.005}(\frac{m_e}{\Lambda_{QCD}})^{0.5}
\end{equation}
Comparison of the Cs hyperfine standard with $SF_6$ molecular vibration
frequencies
was discussed in Ref.~\cite{Chardonnet}. In this case
$X(Cs/Vibrations)= \alpha^{2.8}[m_e/\Lambda_{QCD}]^{0.5}
[m_q/\Lambda_{QCD}]^{0.091}(\frac{m_s}{\Lambda_{QCD}})^{0.011}$.
 
The measurements of hyperfine constant ratios in different isotopes
of the same atom depends on the ratio of magnetic moments and 
is therefore sensitive to $m_q/\Lambda_{QCD}$. For example,
it would be interesting to measure the rate of change for
hydrogen/deuterium ratio where $X(H/D)=[m_q/\Lambda_{QCD}]^{-0.068}
[m_s/\Lambda_{QCD}]^{0.032}$. 

Walsworth has suggested that one might 
measure the ratio of the Zeeman transition
frequencies in noble gases in order to explore the time dependence of the ratio
of nuclear magnetic moments. Consider, for example $^{129}$Xe/$^3$He.
{}For $^3$He the magnetic moment is very close
to that of neutron. For other noble gases the nuclear magnetic
moment is also given by the valence neutron, however, there are significant
many-body corrections. For $^{129}$Xe the valence neutron is
in an $s_{1/2}$ state, which corresponds to the single-particle value
of the nuclear magnetic moment, $\mu=\mu _n= -1.913$. The measured value
is $\mu= -0.778$.
The magnetic moment of the nucleus changes most efficiently
through the spin-spin interaction, because the valence neutron 
transfers a part of its spin, $<s_z>$, to the core protons and 
the proton magnetic
moment is large and has the opposite sign. In this approximation 
$\mu=(1-b)\mu_n + b \mu_p$. This gives b=0.24 and 
the ratio of magnetic moments $Y\equiv \mu (^{129}$Xe)/$\mu (^3$He)$ \approx
0.76 + 0.24 g_p/g_n$. Using 
Eqs.~(\ref{gpsq},\ref{gpss},\ref{gnsq},\ref{gnss}) we obtain
an estimate for the relative variation of
$\mu (^{129}$Xe)/$\mu (^3$He), which can be presented as variation of
$X=[m_q/\Lambda_{QCD}]^{-0.027}[m_s/\Lambda_{QCD}]^{0.012}$.
Here again $\delta Y/Y=\delta X/X$.   

Note that the accuracy of the results presented in this paper
depends strongly on the fundamental constant under study.
The accuracy for the dependence on $\alpha$ is a few percent.
The accuracy for $m_q/\Lambda_{QCD}$ is about $30\%$ -- being limited
mainly by the accuracy of the single-particle approximation for nuclear
magnetic moments. (For comparison, the estimated systematic error
associated with the calculation of the effect of the quark mass
variation is less than 10\%.) Finally, we stress that the relation   
(\ref{mQCD}) between the variation of $\alpha$ and  $m/\Lambda_{QCD}$
has been used solely for purposes of illustration.

\acknowledgments

V.F. is grateful to C. Chardonnet,
S. Karshenboim and R. Walsworth for valuable discussions
and to the Institute
for Advanced Study and the Monell foundation for hospitality
and support.
This work is supported by the Australian Research
Council. 


\appendix

\section{Magnetic Moments}
As explained in the text, we explicitly include the processes shown in
Fig.~\ref{fig:1}, which give rise to the leading and next-to-leading 
nonanalytic behaviour as a function
of quark mass.
\begin{table}
\caption{Chiral coefficients for various diagrams contributing to
  proton and neutron magnetic moments. We use $SU(6)$ symmetry to relate
  the meson couplings to the $\pi N\Delta$ vertex,   
  $\chiC=-2 D$.
\label{tab:chi}}
\begin{tabular}{lcc}
\hline\hline
$\alpha$ & $\beta_{\mu\alpha}^p$ & $\beta_{\mu\alpha}^n$\\
\hline
(a)  & $-(F+D)^2$ 
     & $ (F+D)^2$ \\
(b)  & $-\frac{2}{9}\chiC^2$
     & $ \frac{2}{9}\chiC^2$\\
(c)  & $-\frac{1}{6}(D+3F)^2$
     & $ 0 $\\
(d)  & $-\frac{1}{2}(D-F)^2$
     & $           -(D-F)^2$\\
\hline\hline
\end{tabular}
\end{table}

We describe the quark mass dependence of the magnetic moments as:
\begin{eqnarray}
\mu &=& \frac{\alpha_0}{1+\alpha_2 \mpi^2}+ M^{\rm L}\, , 
\label{eq:mu}
\end{eqnarray}
where $M^{\rm L}$ denotes the chiral loop corrections
given by
\begin{eqnarray}
M^{\rm L}   &=& \chi_{\mu(a)}\,I_{\mu}(\mpi,0,\Lambda) \non\\
            &&  + \chi_{\mu(b)}\,I_{\mu}(\mpi,\Delta_{N\Delta},\Lambda)\non\\
            &&+ \chi_{\mu(c)}\,I_{\mu}(\mK,\Delta_{N\Lambda},\Lambda)\non \\
            &&+ \chi_{\mu(d)}\,I_{\mu}(\mK ,\Delta_{N\Sigma},\Lambda)\, .
\end{eqnarray}
The chiral coefficients of the loop integrals, $\chi_{\mu\alpha}$, are given by
\begin{equation}
\chi_{\mu\alpha} = \beta_{\mu\alpha} \frac{M_N}{8\,\pi\,\fpi^2}
\end{equation}
and are summarised in Table~\ref{tab:chi}
\cite{Theberge:1981mq,Jenkins:1992pi,Leinweber:2002qb}.
Note that the required analytic terms in the chiral expansion to this
order have been placed in a Pad\'e approximant designed to reproduce the
Dirac moment behaviour of the nucleon at moderate quark mass.

The corresponding loop integral is given by
\begin{eqnarray}
I_{\mu}(m,\Delta,\Lambda) = -\frac{4}{3\pi} 
    \int_0^\infty dk\, \frac{(\Delta+2 \wk)k^4u^2(k,\Lambda)}{2\wk^3(\Delta+\wk)^2}
\end{eqnarray}
where the various terms have been defined in Sect. II. 
We note that in the limit
where the mass-splitting vanishes this integral is normalised such that
the leading nonanalytic contribution is $m$.

With the coefficients of the loop integrals defined,  
we only require determination of the parameters
$\alpha_0$ and $\alpha_2$ in Eq.~(\ref{eq:mu}) to constrain the
variation with quark mass. We note also that this form assumes no
analytic dependence on the strange quark mass, beyond what is
implicitly included in the loop diagrams $(c,d)$.  We determine
$\alpha_{0,2}$ for both the proton and neutron by fitting the
physical magnetic moment as well as the lattice QCD data.  
We fit only to the two
heaviest simulation results of the CSSM Lattice Collaboration
\cite{Zanotti:2003gc}, $\mpi^2\sim 0.6$--$0.7\gev^2$. 
These simulations were 
performed with the FLIC fermion action \cite{Zanotti:2001yb} on a
$20^3\times 40$ lattice at $a=0.128\fm$. We select the heaviest two
data points, where the effects of quenching are anticipated to be
small~\cite{Young:2002cj,Young:2003ns}.

The best fits to the physical values and the lattice data give
\begin{eqnarray}
&\alpha_0^p =  2.17\ \NM \hspace*{5mm} \alpha_2^p = 0.817\ \gev^{-2} \\
&\alpha_0^n = -1.33\ \NM \hspace*{5mm} \alpha_2^n = 0.758\ \gev^{-2}\, .
\end{eqnarray}
Upon renormalisation of the loop diagrams, the resultant magnetic
moments in the SU(2) chiral limit are given by
\begin{equation}
\mu_0^p =  3.48\ \NM \, , \hspace*{5mm} {\rm and} \hspace*{5mm} 
\mu_0^n = -2.58\ \NM \, .  \\
\end{equation}

We now take derivatives of Eq.~(\ref{eq:mu}) at the physical pion mass
to determine the variation with quark mass. In particular, we have
\begin{equation}
\delval{\mu} = \left\{ \frac{\mpi^2}{\mu} \frac{d\,\mu}{d\,\mpi^2}
  \right\} \delval{m_q} 
\end{equation}
\begin{equation}
\delval{\mu} = \left\{ \frac{\mKchi^2}{\mu} \frac{d\,\mu}{d\,\mKchi^2}
  \right\} \delval{m_s}\, .
\end{equation}
This yields the results shown in the text.


\end{document}